\documentclass{JHEP3}
\usepackage{axodraw}

\author{ G\"{o}sta Gustafson\\
  Dept.~of Astronomy and Theoretical Physics, Lund University, \\
  Sölvegatan 14A, Lund, Sweden \vspace{2mm}\\
  E-mail: gosta.gustafson@thep.lu.se}

\abstract{In this note we analyse the relation between the triple-pomeron and 
Good--Walker
formalisms for diffractive excitation in DIS and hadronic collisions. In both
approaches gap events are interpreted as the shadow of absorption into
inelastic channels. We here argue that the two formalisms are just different 
views of the same phenomenon. We first demonstrate how this relation works in a
simple toy model, and then show how the relevant features of the toy model are
also realized in real perturbative QCD.}

\title{The Relation between the Good--Walker and Triple-Regge Formalisms for
  Diffractive Excitation}


\preprint{LU-TP 12-21\\
June 2012\\
revised Nov. 2012}
\begin{document}

\section{Introduction}

Events with rapidity gaps contribute of the
order 10\% in DIS at \textsc{Hera} \cite{Adloff:1997sc, Chekanov:2005vv}, 
and in $p\bar{p}$ 
collisions at the CERN $Sp\bar{p}S$ collider \cite{Bernard:1985kh,
Ansorge:1986xq} and the Tevatron \cite{Abe:1993wu}. 
Such events have commonly been interpreted as the shadow of absorption into 
inelastic channels, in analogy with diffraction in optics. Two different
formalisms are based on this analogy: the Mueller--Regge approach 
\cite{Mueller:1970fa, Detar:1971gn} and the
Good--Walker formalism \cite{Good:1960ba}.

In the Regge formalism the absorption is represented by
cut pomeron or reggeon diagrams. Regge diagrams, cut in between two 
exchanged pomerons or reggeons, represent elastic scattering, while
the amplitude for diffractive excitation is determined by the 
$3\rightarrow 3$ scattering process.
At high energies multi-regge diagrams are important,  where the relative 
contributions from cut and uncut diagrams are given by the AGK cutting rules
\cite{Abramovsky:1973fm}. 
Recent analyses within this formalism include work by Kaidalov and
coworkers \cite{Kaidalov:2009hn}, Ostapchenko \cite{Ostapchenko:2010vb}, 
the Durham \cite{Ryskin:2011qe, Ryskin:2012ry}, and the Tel Aviv
groups \cite{Gotsman:2010nw, Gotsman:2012rq, Gotsman:2012rm}.

In the Good--Walker formalism the projectile is written
as a coherent sum of components, which are eigenstates to the interaction with
the target. If the interaction strengths are different for the different
components, the diffractive state no longer corresponds to the incoming coherent
sum, and the cross section for diffractive excitation is determined by the
fluctuations in the interaction strengths. In the analyses mentioned above,
this formalism was used to describe low mass diffraction, but it has also
been proposed to describe the full spectrum of excitation masses in the 
early work by
Miettinnen and Pumplin \cite{Miettinen:1978jb}, and more recently by Hatta
\emph{et al.} \cite{Hatta:2006hs}, and by the Lund group \cite{Avsar:2007xg,
  Flensburg:2010kq, Flensburg:2012zy}. 

The aim of the present note is to argue that, although apparently quite
different, the triple-pomeron and Good--Walker formalisms are just different
views of the same phenomenon. Besides the common basis in the optical theorem,
a common underlying dynamics for the two formalisms
is suggested by the connection between the fluctuations in the cascade 
evolution and the triple-pomeron vertex, which is demonstrated in 
ref.~\cite{Iancu:2005nj}. Within the dipole cascade formalism it is there 
shown that both these effects are determined by the splitting of a dipole
into two connected dipoles, sharing a common gluon
(see also ref.~\cite{Braun:1997nu}). Finally such a relation is also 
supported by the observation in
ref.~\cite{Flensburg:2010kq}, that the result in the Good--Walker formalism
indeed has the expected triple-pomeron form.
%, corresponding to a simple
%pomeron pole and an approximately constant triple-pomeron coupling.

In order to illuminate the relation between the two formalisms, we first study 
a simple toy model. We will later discuss the relation between the toy model and
the real QCD evolution.
The toy model was proposed by Mueller to illustrate effects of multiple 
scattering and saturation \cite{Mueller:1994gb}, but here we will instead use 
it in the weak interaction limit, to demonstrate the similarities between the
triple--pomeron and Good--Walker formalisms for high-mass diffraction. 
The model has the same basic structure as onium-onium
scattering in 4-dimensional QCD, but is made simpler by eliminating the
transverse degrees of freedom. Thus, although it is 
not an approximation to QCD in 4 dimensions, it does 
have the same basic structure as the BFKL pomeron.

Also other approaches to a description of gap events have been presented.
In the ``color reconnection'' model it is assumed that the color
exchange in a ``normal'' inelastic event can be neutralized through the
mediation of soft gluons \cite{Edin:1995gi, Pasechnik:2010cm}, and in the
Ingelman--Schlein picture events in diffractive excitation are
described in terms of a pomeron flux factor and parton distribution functions
in the pomeron \cite{Ingelman:1984ns}. For these approaches the relation to 
diffraction in optics is not clear, and they will not be discussed here.

\section{The toy model}

The toy model has a structure with essential similarities to BFKL evolution in
the dipole formulation of QCD. 
In the model there is only one type of dipole (or parton), which evolves
into a cascade when the rapidity $y$ is increased. A dipole can emit a new
dipole with a probability $\alpha$ per unit rapidity. We let $P_n(y)$
denote the probability for the cascade to contain $n$ dipoles at rapidity
$y$. This distribution satisfies the following evolution equation
\begin{equation}
\frac{d P_n}{d y}= \alpha [(n-1)\,P_{n-1} - n\,P_n].
\end{equation}
Here the first term describes the gain when a new dipole is emitted in a state 
with $(n-1)$ dipoles, and the second term the loss when a dipole emission
changes a state with $n$ dipoles into a state with $(n+1)$ dipoles.
With the boundary value $P_n(0)=\delta_{n1}$, the solution to this
equation is
\begin{equation}
P_n(y)=e^{-\alpha y}(1-e^{-\alpha y})^{n-1}.
\end{equation}
The average and variance of this multiplicity distribution are given by
\begin{eqnarray}
\langle n \rangle &=& e^{\alpha y} \label{eq:averagen}\\
V &\equiv & \langle n^2 \rangle - \langle n \rangle^2 = e^{2\alpha y} -
e^{\alpha y} \label{eq:variance}.
\end{eqnarray}
We note that for asymptotic energies (large $y$) the distribution satisfies
KNO scaling
\begin{equation}
P_n\approx \frac{1}{\langle n\rangle}\, F(n/\langle n\rangle),\,\,\,\mathrm{with}\,\,\,F(\xi)=e^{-\xi}.
\end{equation}
This approximation would give $V=e^{2\alpha y}$, and the second term in
eq.~(\ref{eq:variance}) is a result of the deviation from exact scaling for
finite (and discrete) $n$-values.

When two dipoles meet, they interact inelastically with probability $2\alpha^2
f$, and in the weak interaction limit the optical theorem gives the elastic
dipole-dipole scattering amplitude $\alpha^2 f$. I have here used the same
notation as in ref.~\cite{Mueller:1994gb}. In the QCD analogy $\alpha \propto
\bar{\alpha}_s\equiv N_c \alpha_s/\pi$ and $f\propto 1/N_c^2$.

We study two cascades, evolved in opposite directions distances $y_1$ and
$y_2$ respectively. When they meet the colliding
dipoles interact independently, and in the Born approximation the elastic 
scattering amplitude is given by
\begin{equation}
T_{\mathrm{el}} = \sum_{n,m} P_n(y_1) P_m(y_2) n m\, \alpha^2 f
= \alpha^2 f e^{\alpha y_1} e^{\alpha y_2} = \alpha^2 f e^{\alpha Y}.
\label{eq:tel}
\end{equation}
Here $Y=y_1+y_2$ denotes the total rapidity range between the projectile
and the target, and we note that the result depends only on this total range, 
and thus
is independent of the Lorentz frame used for the calculation. In appropriate 
units we define $s=e^{Y}$, and thus we have in the weak interaction limit
\begin{eqnarray}
\sigma_{\mathrm{el}}=(\alpha^2 f)^2 s^{2\alpha} \label{eq:TRel}\\
\sigma_{\mathrm{inel}}=2\alpha^2 f s^\alpha,
\label{eq:TRtot}
\end{eqnarray}
corresponding to a pomeron with intercept $1+\alpha$, and a dipole-pomeron
coupling $g_{dP}^2=\alpha^2 f$.

\section{Triple-pomeron formalism}

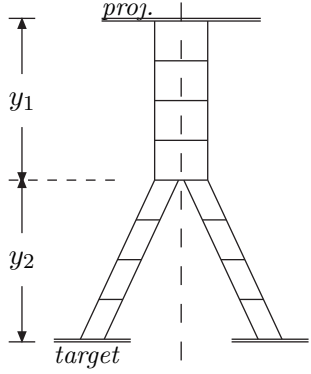
\begin{figure}
\begin{center}
\begin{picture}(150,130)(-20,5)
\Text(20,131)[bl]{{\small \textit {proj.}}}
\Line(20,130)(80,130)
\Line(20,131)(80,131)
\Line(40,130)(40,70)
\Line(60,130)(60,70)
\Line(40,115)(60,115)
\Line(40,100)(60,100)
\Line(40,85)(60,85)
\Line(40,70)(60,70)

%\Text(90,100)[l]{{\large $\Psi_{proj}=\sum_n c_n \Phi_{p,n}$}}

\Line(40,70)(12,10)
\Line(49,70)(21,10)
\Line(51,70)(79,10)
\Line(60,70)(88,10)
\Line(33,55)(42,55)
\Line(58,55)(67,55)
\Line(26,40)(35,40)
\Line(65,40)(74,40)
\Line(19,25)(28,25)
\Line(72,25)(81,25)
\Line(2,10)(31,10)
\Line(2,9)(31,9)
\Line(69,10)(98,10)
\Line(69,9)(98,9)
\Text(2,7)[tl]{{\small \textit {target}}}

\Line(-15,131)(-5,131)

\Line(-15,9)(-5,9)
\LongArrow(-10,110)(-10,130)
\LongArrow(-10,90)(-10,70)
\LongArrow(-10,50)(-10,70)
\LongArrow(-10,30)(-10,10)
\Text(-10,100)[]{$y_1$}
\Text(-10,40)[]{$y_2$}
\DashLine(-15,70)(25,70){4}
%\SetColor{\PSBrlu}
\DashLine(50,137)(50,2){7}
\end{picture}
\end{center}
\caption{\label{fig:GW}A triple-pomeron diagram}
\end{figure}

In the triple-pomeron formalism for diffractive excitation the amplitude is
determined by the $3\rightarrow 3$ scattering process. We study a
situation where a parent dipole evolves a distance $y_1$, when one of its
$n$ daughter dipoles is split in two. These two dipoles interact
elastically (and independently) with two separate dipole cascades, both 
developed a distance $y_2$ (see fig.~\ref{fig:GW}). 
The probability for such a split per unit rapidity is
$\alpha$, and in the weak interaction limit the interaction weight for
each of the two interactions is $\alpha^2 f$. With $M_X^2=e^{y_1}$ and
$s=e^{y_1+y_2}$ this gives the cross section for single diffractive
excitation
\begin{eqnarray}
M_X^2 \frac{d \sigma_{\mathrm{SD}}}{d M_X^2}=\alpha
\sum_n P_n(y_1) n \sum_m P_m(y_2) m\, (\alpha^2 f) \sum_l P_l(y_2) l\,
(\alpha^2 f)=\nonumber \\
= \alpha^5 f^2 e^{\alpha y_1} (e^{\alpha y_2})^2
=\alpha^5 f^2 (M_X^2)^\alpha (s/M_X^2)^{2\alpha}.
\label{eq:tripleregge}
\end{eqnarray}
This result corresponds to the triple-pomeron expression, where the weight factor
$\alpha^5 f^2$ should correspond to $g_{dP}^3 g_{3P}$ (with a
suitably defined triple-pomeron coupling $g_{3P}$). With 
$g_{dP}^2=\alpha^2 f$ this gives $g_{3P}=\alpha^2\sqrt{f}$,
where a factor $\alpha\sqrt{f}$ ($\sim \alpha_s$) comes from the interaction 
weights, and a factor $\alpha$ ($\sim \bar{\alpha}_s$) from the splitting 
probability per unit rapidity.

We can in the same way also calculate the total diffractive cross section
by replacing the factor $n$ in eq.~(\ref{eq:tripleregge}) by $n^2$. This
will include contributions from interaction of two uncorrelated dipoles in
the projectile cascade, which corresponds to all possible masses 
$M_X^2<M_{X,\mathrm{max}}^2=e^{y_1}$
(including elastic scattering). The result can be written in the form
\begin{eqnarray}
\sigma_{\mathrm{tot\,proj\,diff}}=
\sum_n P_n(y_1) n^2 \sum_m P_m(y_2) m\, (\alpha^2 f) \sum_l P_l(y_2) l\,
(\alpha^2 f)= \nonumber \\
= \alpha^4 f^2 (2e^{2\alpha y_1}-e^{\alpha y_1}) \, e^{2\alpha
y_2}
=\alpha^4 f^2 \left[(s)^{2\alpha} + (s)^{2\alpha}(1-
  1/(M_{X,\mathrm{max}}^2)^{\alpha})\right]. 
\label{eq:triplereggeincl}
\end{eqnarray}
Here the first term in the square parenthesis corresponds to the elastic
cross section from eq.~(\ref{eq:TRel}). The second term represents the
integrated sum over diffractive masses in eq.~(\ref{eq:tripleregge}), given by
the single diffractive excitation cross section
\begin{equation}
\sigma_{\mathrm{SD}}= \int_0^{y_1} dy_1 \alpha^5 f^2 e^{2\alpha Y}
e^{-\alpha y_1}=\alpha^4 f^2 e^{2\alpha Y}(1-e^{-\alpha y_1})
\label{eq:SDtr}
\end{equation}

Subtracting the elastic cross section, given by the square of the amplitude in
eq.~(\ref{eq:tel}), from the result in eq.~(\ref{eq:triplereggeincl}), we note
that the diffractive excitation also can be written in the form 
\begin{equation}
\sigma_{\mathrm{SD}}= \left\{\sum_n P_n(y_1) n^2 -(P_n(y_1) n)^2 \right\}
\left(\sum_m P_m(y_2) m\, (\alpha^2 f)\right)^2.
% \sum_l P_l(y_2) l\,(\alpha^2 f)
\label{eq:SDfluc}
\end{equation}
Thus we see that the result is actually \emph{determined by the fluctuations} 
in the multiplicity distribution $P_n(y_1)$, given by the variance 
$\langle n^2\rangle -\langle n\rangle^2$.

\section{Good--Walker formalism}

In the Good--Walker formalism the mass eigenstates $\Psi_k$
are written as linear combinations of the diffractive
eigenstates $\Phi_i$, $\Psi_k=\sum c_{ki}\Phi_i$, where $c_{ki}$ is a unitary 
matrix. For scattering against a structureless target, and with $\Psi_{\mathrm{in}}=\Psi_0$, the elastic amplitude is given by the average over the diffractive eigenstates:
\begin{equation} 
\langle \Psi_0 | T | \Psi_0 \rangle = \sum |c_{0i}|^2 T_i 
= \langle T \rangle.
\label{eq:Tel}
\end{equation}
The amplitude for diffractive transition to the mass eigenstate $\Psi_k$
becomes
\begin{equation}
\langle \Psi_{k} | T | \Psi_0 \rangle = \sum_i  c^*_{ik} T_i c_{0i}.
\label{eq:Tik}
\end{equation}
Neglecting the transverse coordinates this gives a total diffractive cross 
section
\begin{equation}
\sigma_{\mathrm{diff}}
=\sum_k \langle \Psi_0 | T | \Psi_{k} \rangle \langle \Psi_{k} | T |
\Psi_0 \rangle =\langle T^2 \rangle,
\label{eq:GWsigmadiff}
\end{equation}
and after subtracting the elastic scattering, the cross section for 
diffractive excitation is given by the \emph{fluctuations}:
\begin{equation}
\sigma_{\mathrm{diff \,exc}}  = \langle T^2 \rangle - \langle T \rangle ^2\equiv V_T,
\label{eq:eikonaldiff}
\end{equation}

For scattering of two dipole cascades, the cross section for diffractive
scattering of the projectile (including elastic), and elastic scattering of 
the target, is given by
$ \langle \langle T \rangle_{\mathrm{targ}}^2
\rangle_{\mathrm{proj}} $. 
In the toy model the cross section, $\sigma_{SD}$, for diffractive excitation of
the \emph{projectile}, and elastic scattering of the \emph{target}, 
is thus given by
\begin{eqnarray}
\sigma_{\mathrm{SD}}&=&\sigma_{\mathrm{tot\,proj\,diff}} - \sigma_{\mathrm{el}} = \langle \langle T \rangle_{\mathrm{targ}}^2
\rangle_{\mathrm{proj}} 
-(\langle \langle T \rangle_{\mathrm{targ}}\rangle_{\mathrm{proj}})^2=
\nonumber \\
&=&\sum_n P_n(y_1) n^2 \Bigl(\sum_m P_m(y_2) m \alpha^2 f\Bigr)^2 -
\Bigl(\sum_n P_n(y_1) n \sum_m P_m(y_2) m \alpha^2 f\Bigr)^2  
= \nonumber \\
&=&%(\alpha^2 f)^2 (e^{2\alpha y_1} - e^{\alpha y_1})(e^{2\alpha y_2})=
(\alpha^2 f)^2 e^{2\alpha Y}(1 - e^{-\alpha y_1}).
\label{eq:GWSD}
\end{eqnarray}
This cross section includes all excited systems $X$ confined within the
rapidity range $0<y<y_1$, and the average over target and projectile states 
correspond to $\sum_m P_m$ and $\sum_n P_n$ respectively.
We see that the result exactly reproduces the 
triple-pomeron result in eqs.~(\ref{eq:SDfluc}, \ref{eq:SDtr}); in both
formalisms the cross section is determined by the fluctuations in the 
evolution. Taking the derivative with respect to $y_1$,
we get also in the Good--Walker formalism the differential result in 
eq.~(\ref{eq:tripleregge}).

\section{Relation to QCD}

We have seen above that in the toy model the dipole splitting probability 
per unit rapidity,
$\alpha$,  determines the fluctuations in the evolution in 
eq.~(\ref{eq:variance}), and also the differential diffractive cross section
in eq.~(\ref{eq:tripleregge}) via the factor $\alpha \sum_n P_n(y_1) n$.
Analogous results are obtained for real QCD in three space dimensions by Iancu 
and Triantafyllopoulos in~\cite{Iancu:2005nj}, which show that the dipole
splitting term determines both the \emph{fluctuations in the evolution} and the
\emph{triple-pomeron coupling}. In their notation the probability per unit
rapidity for a dipole $(\mathbf{x,y})$ to split into two connected dipoles,
$(\mathbf{x,z})$ and $(\mathbf{z,y})$, (sharing a common gluon) is given by 
\begin{equation}
\frac{\bar{\alpha}_s}{2\pi}
\mathcal{M}(\mathbf{x,y,z})=\frac{\bar{\alpha}_s}{2\pi} 
\frac{(\mathbf{x-y})^2}{(\mathbf{x-z})^2 (\mathbf{z-y})^2}.
\label{eq:split}
\end{equation}
The dipole density after evolution $y$ units of rapidity, is denoted
$n_y(\mathbf{x,y})$, and the dipole-dipole scattering cross section
$\mathcal{A}_0(\mathbf{x,y}|\mathbf{u,v})$. With these notations the
differential diffractive cross section can be extracted from their
eq.~(5.5), and be expressed in the following form (the integral sign indicates
appropriate integrations over transverse coordinates)
\begin{equation}
M_X^2 \frac{d \sigma_{\mathrm{SD}}}{d M_X^2}\propto 
\int \bar{\alpha}_s\mathcal{M}(\mathbf{u,v,z}) n_{y_1}(\mathbf{u,v})
n_{y_2}(\mathbf{x_1,y_1})\alpha_s^2\mathcal{A}_0(\mathbf{x_1,y_1}|\mathbf{u,z})
n_{y_2}(\mathbf{x_2,y_2})\alpha_s^2\mathcal{A}_0(\mathbf{x_2,y_2}|\mathbf{z,v}).
\label{eq:SDIT}
\end{equation}
With the translation $\bar{\alpha}_s\mathcal{M}\rightarrow \alpha$ and 
$\alpha_s^2\mathcal{A}_0 \rightarrow \alpha^2 f$, this result exactly
corresponds to the toy model expression in eq.~(\ref{eq:tripleregge}).
Thus we conclude that the relation between the triple-pomeron and Good--Walker
formalisms should hold also in real QCD.
Important for the result is also the fact, that these fluctuations are large,
and lead to approximate KNO scaling, which is a basic feature of QCD evolution
(see \emph{e.g.} ref.~\cite{Malaza:1985jd}).

It ought to be noticed that  
in LL BFKL evolution the pomeron singularity is a cut, which implies that the
triple-pomeron coupling obtained from the integral over $\mathcal{M}$ in
eq.~(\ref{eq:SDIT}) will depend on the energy. It is, however,
well known that with a running coupling the cut is replaced by a series
of poles \cite{Lipatov:1985uk}. For high energies the leading pole is 
dominating, and if
the corresponding eigenfunction to the evolution operator is denoted
$\phi(\mathbf{x-y})$, the triple-pomeron coupling is given by the energy
independent expression
\begin{equation}
g_{3P} \propto \int \phi(\mathbf{x-z}) \phi(\mathbf{z-y}) \bar{\alpha}
\mathcal{M}(\mathbf{x,y,z}) \phi(\mathbf{x-y}) d^2 x d^2 z.
\end{equation}

The connection between the triple-pomeron and Good--Walker formalisms is also
reflected in the results from the dipole cascade model DIPSY,
which is based on BFKL evolution and the Good--Walker picture for diffractive
excitation. 
These results indeed reproduce the expected triple-pomeron form, with a simple
pomeron pole and an approximately constant
triple-pomeron coupling, once the non-linear
effects are switched off~\cite{Flensburg:2010kq}.\footnote{In 
  perturbative QCD with massless gluons and without confinement, the 
  triple-pomeron
  coupling has a singularity $\sim 1/\sqrt{-t}$ \cite{Mueller:1994jq, 
  Bartels:2002au}. In the DIPSY model confinement effects are also included 
  (they are also important for satisfying the Froissart--Martin bound at 
  asymptotic energies~\cite{Avsar:2008dn}). Also in recent triple-regge fits by
  the Tel Aviv \cite{Gotsman:2012rq} and Durham \cite{Ryskin:2012ry} groups, 
  the results correspond to a simple pomeron pole.}

\section{Discussion}

Essential for the results described above is that in both the triple-pomeron
and Good--Walker formalisms, the diffractive excitation is determined by the
fluctuations.  We also note that it is the small 
correction term in the expression for the
variance in eq.~(\ref{eq:variance}), which is responsible for the differential
diffractive cross section in eq.~(\ref{eq:tripleregge}). At high energies
(and disregarding saturation effects) 
the integrated single diffractive cross section saturates at $\sigma_{\mathrm{SD}}
=(\alpha^2 f)^2 e^{2\alpha Y}$, given by the scaling part in the variance 
$\propto e^{2\alpha
y}$ in eq.~(\ref{eq:variance}). From the fact that $d\sigma_{\mathrm{SD}}/d
\ln(M_X^2)$ drops for large $M_X$ at constant $s$, one could possibly get the
impression that the fluctuations are less important in this case. This
is, however, not true. For a fixed target system with constant $y_2$, the
single diffractive cross section would grow $\sim e^{2\alpha
  y_1}=(M_{X,\mathrm{max}}^2)^{2\alpha}$.  
For fixed $s=e^{(y_1+y_2)}$ the increased fluctuations in the projectile
cascade for larger $y_1$ ($\propto (e^{2\alpha y_1}-e^{\alpha y_1})$)
are compensated by an approximately equally fast decrease in the 
target evolution $\propto (e^{\alpha y_2})^2$. The suppressed term $-e^{\alpha
y_1}$ in the variance in eq.~(\ref{eq:variance}) implies that the differential
diffractive excitation $d\sigma_{\mathrm{SD}}/d y_1$ 
is proportional to $\alpha e^{-\alpha y_1}=\alpha/(M_X^2)^\alpha$, over 
a range of the order $1/\alpha$ in $y_1=\ln(M_X^2)$.  

The Good--Walker formalism has the advantage that it is highly predictive.
Including saturation effects (related to enhanced and
semienhanced diagrams) the DIPSY model, based on the Good--Walker formalism, 
reproduces experimental results for diffraction in DIS and $pp$ collisions,
without extra tunable parameters besides those tuned to the total and elastic
cross sections~\cite{Avsar:2007xg, Flensburg:2010kq, Flensburg:2012zy}. 
Thus \emph{e.g.} the triple-pomeron coupling is fully 
determined by the dynamics via the optical theorem. 
This is in contrast to phenomenological applications of the 
triple-regge formalism, where the
triple-pomeron coupling is an adjustable parameter, which has to be fitted to
data. 

In the discussions above we have only discussed pomeron exchange, which
dominates for high energies and excitation masses,
$s>>M_X^2>>\Lambda_{\mathrm{QCD}}$. For lower energy or excitation mass, also
lower regge
trajectories contribute. These contributions are not included in the DIPSY 
dipole model. Nevertheless, with a cutoff for low masses, $M_X$, of the order of
1 GeV, the results are consistent with data from
refs.~\cite{Adloff:1997sc}-\cite{Abe:1993wu}. (As far as we know, no model
is able to reproduce the full resonance structure of low mass excitations.)

\section{Conclusion} 

In conclusion we argue that the dynamics of the BFKL pomeron implies that the
triple-pomeron and the Good--Walker formalisms for high mass diffraction 
are just different formulations of the same phenomenon. In both formalisms
diffractive excitation is the shadow of absorption into inelastic channels, and
the similarity is related to the KNO scaling in QCD
evolutions, where the fluctuations in the gluon multiplicity grow proportional 
to the multiplicity, $\sigma^2 \approx \langle n \rangle^2$.  A
phenomenological analysis based on the Good--Walker formalism has the
advantage that diffractive excitation is fully determined by the dynamics via
the optical theorem, with no extra parameters beyond those tuned to the total
and elastic cross sections.

The relation between the two formalisms is here illustrated by an analysis of
a simple toy model in one 
plus one dimensions, but we also see that the essential features of this model
are shared by BFKL evolution in real QCD in three space dimensions.
The relation is further supported by results from the Lund dipole cascade
model DIPSY, based on BFKL evolution and using the Good--Walker formalism. 
Excluding non-linear
effects, the model reproduces the triple-pomeron form for diffractive excitation
\cite{Flensburg:2010kq}, and including also effects of saturation it reproduced
both inclusive and exclusive diffractive data in DIS and $pp$ 
collisions~\cite{Avsar:2007xg, Flensburg:2010kq, Flensburg:2012zy}.

\section*{Acknowledgments} 

The author is grateful to Leif L\"{o}nnblad for valuable
discussions. I also want to thank the referee for pointing out the resemblence
between the toy model and the results in 
refs.~\cite{Iancu:2005nj, Braun:1997nu}. 

\bibliographystyle{utcaps}
\bibliography{refs}

\end{document}